\documentclass[epj]{svjour}
\usepackage{graphics}
\begin{document}
\title{Fundamental Symmetries and Interactions - Some Aspects} 
\author{Klaus Jungmann\inst{1} 
}                     
\institute{Kernfysisch Versneller Instituut, Zernikelaan 25, 9747 AA Groningen, The Netherlands}
\date{Received: date / Revised version: date}
%
\abstract{
In the framework of nuclear physics and at nuclear physics facilities
a large number of different experiments can be performed which render
the possibility to investigate fundamental symmetries and interactions
in nature.
In particular, the precise measurements of 
properties of fundamental fermions,
searches for new interactions in $\beta$-decays,
and violations of discrete symmetries have a robust discovery 
potential for physics beyond standard theory.
Precise measurements of fundamental constants
can be carried out as well. Low energy experiments  
allow probing of New Physics models 
at mass scales far beyond the reach of
present accelerators or such planned  for the future
in the domain of high energy physics 
and at which predicted new particles could be produced directly.
} 
\maketitle
\section{Introduction}
\label{intro}
Symmetries play an important and central role in physics. Where\-as
global symmetries relate to conservation laws, local symmetries 
yield forces \cite{Lee_56}. Today four fundamental interactions are
known in physics:
(i)    Electromagnetism, 
(ii)     Weak Interactions,
(iii)    Strong Interactions, and 
(iv)   Gravitation.
These forces are considered fundamental, because all
observed dynamic processes in nature can be traced back to one or
a combination of them. Together with fundamental symmetries they from 
the framework on which all physical descriptions ultimately rest.

The Standard Model (SM) is a most remarkable theory.
Electromagnetic, Weak and many aspects of Strong 
Interactions can be described to astounding precision in one single 
coherent picture. It is a major goal in modern physics to find a unified 
quantum field theory which includes all the four known
fundamental forces. To achieve this, a satisfactory quantum 
description of gravity remains yet to 
be found. This is a lively field of actual activity. 

In this write-up we are concerned with
important implications of the SM. In particular, 
searches for new,  yet unobserved interactions play a central role.
At present, such are suggested by a variety of speculative models
in which extensions to the present standard theory are introduced
in order to explain some of the features in the SM, which are
not well understood and not well founded, although the 
corresponding experimental facts are accurately described.
Among the intriguing questions in modern physics are 
the number of fundamental particle generations and
the hierarchy of the  fundamental fermion masses.
In addition, the electro-weak SM has a rather large number of some 
27 free parameters, which all need to be extracted from experiments.
It is rather unsatisfactory that
the physical origin of the observed breaking of discrete 
symmetries in weak interactions,
e.g. of parity (P), of time reversal (T) and of 
combined charge conjugation and parity (CP), 
remains unrevealed, although the experimental findings can be well
described within the SM.

The speculative models beyond the present standard theory
include such which involve left the number of fundamental particle generation-right symmetry, 
fundamental fermion compositeness, new particles, leptoquarks, 
supersymmetry, supergravity and many more. Interesting candidates 
for an all encompassing quantum field theory are string or membrane
(M) theories which in their low energy limit may include supersymmetry.
Without secure future experimental evidence all of these speculative
theories will remain without status in physics, independent of
the mathematical elegance and partial appeal. Experimental searches
for predicted unique features of those models are therefore essential
to steer theory towards a better and deeper understanding of 
fundamental laws in nature. 

In the field of fundamental interactions there are
two important lines of activities: Firstly, there are searches for physics beyond the SM in order
to base the description of all physical processes on a conceptually 
more satisfying foundation, and, secondly, the application of solid knowledge
in the SM for extracting fundamental quantities and achieving a description of more 
complex physical systems, such as atomic nuclei. 
Both these central goals can be achieved at upgraded present and novel, yet to be built 
facilities. In this connection a high intensity proton driver would serve to
allow novel and more precise measurements in a large number of
actual and urgent issues.\cite{NUPECC_2004}.

In this article we can only address a few aspects of a rich spectrum of possibilities. 

\section{Fundamental Fermion Properties}

\subsection{Neutrinos}
The SM knows three charged leptons (e$^-$, $\mu^-$, $\tau^-$)   and
three electrically neutral neutrinos  ($\nu_e$, $\nu_{\mu}$, $\nu_{\tau}$)
as well as their respective antiparticles. 
The members of the lepton families do not participate in strong interactions. 
Neutrinos eigenstates of mass  ($\nu_1$, $\nu_2$, $\nu_3$) 
and flavor are  different and connected to each other through a mixing matrix
analogous to the Cabbibo-Kobaya\-shi-Maskawa mixing in the quark sector (see \ref{CKM}).
The reported evidence for neutrino oscillations strongly indicate finite $\nu$ masses.
Among the recent discoveries are the surprisingly large
mixing angles $\Theta_{12}$ and $\Theta_{23}$.
The mixing angle $\Theta_{13}$, the phases for CP-violation,
the question whether $\nu$'s are Dirac or Majorana particles
and a direct measurement of a neutrino mass
rank among the top issues in neutrino physics \cite{neutrino_reviews}. 

\subsubsection{Novel Ideas in the Neutrino Field}
Two new and unconventional neutrino detector ideas have come up and gained support in the 
recent couple of years,
which have a potential to contribute significantly towards solving
major puzzling questions in physics.

The first concept employs the detection
of high energetic charged particles originating from neutrino reactions 
through Cherenkov radiation in the microwave region (or even sound waves), which results, if 
such particles interact with,e.g., the Antarctic ice or the salt in large salt domes
as they can be found also in the middle of Europe \cite{Gorham_2002}. One advantage of 
such a detector is its
larger density as compared to water, the typical detector material used up to date. 
It remains to be verified whether this concept will also be applicable for high energetic
accelerator  neutrinos, if timing information and narrowband radio detection techniques will
be employed.

The second concept allows directional sensitivity for low energy anti-neutrinos.
The reaction $\overline{\nu} + p \rightarrow e^+ + {\rm n}$ has a 1.8 eV threshold. The resulting
neutron (n) carries directional information in its angular distribution after the event. In
typical organic material the neutron has a range r$_{\rm n}$ 
of a few cm. With a detector consisting of tubes with a
diameter of order r$_{\rm n}$ and with, e.g., boronated walls the resulting $\alpha$-particle from the n+B
nuclear reaction can be used to determine on average the direction of incoming anti-neutrinos.
Such a detector, if scaled to sufficient mass, can be used to determine the
distribution of radionuclides in the interior of the earth (including testing the rather exotic
of a nuclear reactor in center of the earth) \cite{deMeijer_2004}.

A further 
rather promising application of such a detector
would be a measurement of the neutrino generation mixing angle $\Theta_{13}$ in  a reactor
experiment with a near and far detector in  $\approx$ few 100~m and  $\approx$ few 100 km 
distance. For this measurement the 
importance of directional sensitivity for low energy $\nu$'s is an indispensable requirement.

\subsubsection{Neutrino Masses}
The best neutrino mass limits have been extracted from measurements of the
tritium $\beta$-decay spectrum close to its endpoint.
Since neutrinos are very light particles, a mass measurement can best 
be performed in this region of the spectrum as in other parts the 
nonlinear dependencies caused by the relativistic nature of the kinematic problem 
cause a significant loss of accuracy. This by far overwhelms the possible gain in statistics
one could hope for. Two groups in Troitzk and Mainz
used spectrometers based on Magnetic  Adiabatic Collimation combined with an Electrostatic filter
(MAC-E technique) and found $m(\nu_e) < 2.2~eV$ \cite{MainzTroitzk,Weinheimer_2003}. 

A new experiment, KATRIN \cite{Osipowicz_2001}, is presently prepared
in Karlsruhe, Germany, which is planned to exploit the same technique. It
aims for an improvement by about one order of magnitude.  The physical dimensions
of a MAC-E device
scale inversely with the possible sensitivity to a finite neutrino mass. 
This  may ultimately limit an approach with this principle. 
The new experiment will be sensitive to the mass range
where a  finite effective neutrino mass value of between 0.1 and 0.9 eV was
extracted from a signal in neutrinoless double $\beta$-decay in $^{76}$Ge 
\cite{Klapdor_2004}. The Heidelberg-Moskow collaboration performing this
experiment in the Grand Sasso laboratory reports a 4.2 standard deviation
effect for the existence of this decay. 
        \footnote{A number of further
        experiments is under way using different candidate nuclei
        to verify this claim. 
        An extensive coverage of this subject is well beyond the scope of this article.} 
It should be noted that neutrinoless double $\beta$-decay is only possible
for Majorana neutrinos. A confirmed signal would solve one of
the most urgent questions in particle physics. 

Additional work is needed  to obtain more accurate values of the 
nuclear matrix elements which determine the 
lifetimes of the possible neutrinoless double $\beta$-decay candidates. Only then a positive
signal could be 
converted in a Majorana neutrino mass with small uncertainties 
\cite{Elliot_2004}.

\subsection{Quarks - Unitarity of Cabbibo-Kobayashi-Mask- awa-Matrix}
\label{CKM}
The mass and weak eigenstates of the six quarks (u,d,s,c,b, t) are different and related to each other
by a $3 \times 3$ unitary matrix, the Cabbibo-Kobayashi-Maskawa (CKM) matrix. Non-unitarity
of this matrix would be an indication of physics beyond the SM and could be
caused by a variety of possibilities, including the existence of more than
three quark generations or yet undiscovered muon decay channels. The unitarity of the  CKM matrix
is therefore a severe check on the validity of the standard theory and
sets bounds on speculative extensions to it. 

The best test of unitarity results from the first row of the 
CKM matrix through
\begin{equation}
|{\rm V}_{ud}|^2 + |{\rm V}_{us}|^2 + |{\rm V}_{ub}|^2 = 1 -\Delta ,
\end{equation}
where the SM predicts $\Delta$ to be zero. The size 
of the known elements determine that with the present uncertainties
only the elements V$_{ud}$ and V$_{us}$ play a role. 
V$_{ud}$ can be extracted with best accuracy from the ft values of 
superallowed $\beta$-decays. Other possibilities are the neutron
decay and the pion $\beta$-decay, which both are presently studied.

V$_{us}$ can be extracted from K decays and in principle also from
hyperon decays.
One of the triumphs of nuclear physics in contributing to a confirmation
of the standard theory had remained covered for a long time by
a remarkable misjudgment on the side of the Particle Data Group  \cite{PDG}.
This expert panel had decided to increase the uncertainty
   of V$_{ud}$ from nuclear $\beta$-decay \cite{Toner_2003} based on their feelings that
   nuclei would be too complicated objects to trust theory.
   Interestingly, their own evaluation of V$_{us}$ based on Particle Data Group fits
   of K-decay branching ratios turned out to be not in accordance
   with recent independent direct measurements. As a result of the earlier
   too optimistic error estimates in this part a large activity to
   test the unitarity of the CKM matrix took off,
   because a between 2 and 3 standard deviation from unitarity 
   had been persistently reported without true basis\cite{Abele_2004}. 
   Recent careful analysis of the  subject 
   has also revealed overlooked inconsistencies in the overall picture
    \cite{Czarnecki_2004,Ellis_2004} and at this time new determinations of
    V$_{us}$ together with V$_{ud}$ from nuclear $\beta$-decay
    confirm  $\Delta=0$ and 
    the unitarity of the CKM matrix up to presently possible accuracy.        
   
   Because of the cleanest and therefore
   most accurate theory pion $\beta$-decay offers for future higher
   precision measurements the best opportunities, in principle.
   The estimate \cite{Hardy_2004} for accuracy improvement from nuclear $\beta$-decays
   is about a factor 2.
   The  main difficulty for new round rests therefore primarily 
   with finding an experimental technique
   to obtain sufficient experimental accuracy for pion $\beta$-decay. 

\section{Discrete Symmetries}

\subsection{Parity}
The observation of neutral currents together with the
observation of parity non-conserva\-tion in atoms were
important to verify the validity of the SM. The fact that 
physics over 10 orders in momentum transfer - from atoms to highest energy scattering -
yields the same electro-weak parameters may be viewed as 
one of the biggest successes in physics to date.%

However, at the level of highest precision electro-weak experiments
questions arose, which ultimately may call for a refinement.     
The predicted running of the weak mixing angle $sin^2 \Theta_W$
appears  not to be in agreement with observations \cite{Czarnecki_1998}.
If the value of   $sin^2 \Theta_W$  is fixed at the Z$^0$-pole, deep inelastic electron scattering
at several GeV appears to yield a considerably higher value. A reported disagreement
from atomic parity violation in Cs  has disappeared after a revision of atomic theory.

A new round of experiments is being started with the Q$_{weak}$ experiment \cite{Qweak} at the 
Jefferson Laboratory in the USA.  For atomic parity violation in principle
higher experimental accuracy will be possible from experiments using
Fr isotopes \cite{Atutov_2003,Gomez_2004}
 or single Ba or Ra ions in radiofrequency traps \cite{Fortson}.
Although the weak effects are larger in these systems due to their high power dependence on
the nuclear charge, this can only be exploited after better atomic wave function
calculations will be available, as the observation is always through 
an interference of weak with electromagnetic effects.

\subsection{Time Reversal and CP Violation}
The role of a violation of combined charge conjugation (C) and parity (P) 
is of particular importance
through its possible relation to the observed matter-antimatter 
asymmetry in the universe.  This connection 
is one of the strong motivations to
search for yet unknown sources of 
CP violation. A. Sakharov \cite{Sakharov_1967} has suggested that the 
observed dominance of matter
could be explained via CP-violation in the early universe 
in a state of thermal non-equilibrium and with baryon number violating
processes. CP violation as described in the SM 
is insufficient to satisfy the needs of this elegant model.
Permanent Electric Dipole Moments (EDMs) certain
correlation observables  in $\beta$-decays offer excellent opportunities 
to find new sources of CP-violation.

\subsubsection{Permanent Electric Dipole Moments (EDMs)}
An EDM of any fundamental particle
violates both parity and time reversal (T) symmetries. 
With the assumption of 
CPT invariance a  permanent dipole moment also violates CP.
EDMs  for all particles are caused by CP violation as it is known from 
the K systems through higher order loops. These are at least 4 orders of magnitude below the
present experimentally established limits. Indeed, a large number of speculative models
foresees permanent electric dipole moments which could be as large as 
the present experimental limits just allow. 
Historically the non-observation of permanent electric dipole moments has
ruled out more speculative models than any other experimental
approach in all of particle physics \cite{Ramsey_1999}. 
EDMs have been
searched for in various systems with different sensitivities 
(Table \ref{EDMs}). In composed systems such as molecules
or atoms fundamental particle dipole moments of constituents may be
significantly enhanced\cite{Sandars_2001}. Particularly in polarizable
systems there can exist large internal fields.

There is no preferred system to search for an EDM. In fact,
many systems need to be examined, because depending
on the underlying process different systems have
in general quite significantly different susceptibility
to acquire an EDM through a particular mechanism.
In fact, one needs to investigate different systems.
An EDM may be found an ''intrinsic property'' of
an elementary particle asw we know them, because the underlying 
mechanism is not accessible at present. However, it can also
arise from CP-odd forces between the constituents under observation,
e.g. between nucleons in nuclei or between nuclei and
electrons. Such EDMs could be much higher than such
expected for elementary particles originating within the popular, usually 
considered standard theory models. No other constraints are known.

This highly active field of research benefited recently from a number of novel 
developments.
One of them concerns the Ra atom, which  
has rather close lying $7s7p^3P_1$ and $7s6d^3D_2$ states.
Because they are of opposite parity, a significant enhancement 
has been predicted  for an  electron EDM \cite{Dzuba_2001}, much
higher than for any other atomic system. Further more,
many Ra isotopes are in a region where (dynamic) octupole deformation occurs
for the nuclei, which also may enhance the effect of a nucleon EDM
substantially, i.e. by some two orders of magnitude \cite{Engel_2004}.
From a technical point of view the Ra atomic levels of interest for en experiment
are well accessible spectroscopically and a variety of isotopes can be produced 
in nuclear reactions. 
The advantage of an accelerator based Ra experiment is apparent,
because EDMs require isotopes with spin
and all Ra isotopes with finite nuclear spin are relatively short-lived
\cite{Jungmann_2002}.

A very novel idea was introduced recently for measuring an 
EDM of charged particles. The high
motional electric field is exploited, which charged particles at relativistic speeds 
experience in a magnetic storage ring.
In such an experiment the Schiff theorem can be circumvented
(which had excluded charged particles from experiments due to the
Lorentz force acceleration) because of the non-trivial geometry of the problem
\cite{Sandars_2001}. With an
additional radial electric field in the storage region the spin precession due to the
magnetic moment anomaly can be compensated, if the 
effective magnetic anomaly $a_{eff}$ is small, i.e. $ a_{eff}<<1$. 
The method was first considered for muons. For longitudinally polarized 
muons injected into the ring an EDM
would express itself 
as a  spin rotation out of the orbital plane.
This can be observed as a time dependent (to first order linear in time) 
change of the above/below  the plane of orbit counting rate ratio. 
For the possible muon beams at the future J-PARC facility in Japan
a sensitivity of $10^{-24}$~e\,cm is expected \cite{Yannis_2003,Farley_2004}. 
In such an experiment the possible muon flux is a major limitation.
For models with nonlinear mass scaling of EDM's such an experiment would 
already be more sensitive to some certain new physics models
than the present limit on the electron EDM 
\cite{Babu_2000}. 
An experiment carried out at a more intense muon source could provide
a significantly more sensitive probe to CP violation in the second 
generation of particles without strangeness \cite{Feng_2004}.

The deuteron is the simplest known nucleus. Here an EDM
could arise not only from a proton or a neutron EDM, but also
from CP-odd nuclear forces \cite{Hisano_2004}. It was shown very recently \cite{Liu_2004} that
the deuteron can be in certain scenarios significantly more sensitive than the
neutron. In equation (\ref{nedm}) this situation is
evident for the case of quark chromo-EDMs:
\begin{eqnarray}
d_{\mathcal{D}} & = & -4.67\, d_{d}^{c}+5.22\, d_{u}^{c}\,,\nonumber \\
d_{n} & = & -0.01\, d_{d}^{c}+0.49\, d_{u}^{c}\,.\label{nedm}
\end{eqnarray}
It should be noted that because of its rather small magnetic anomaly
the deuteron is a particularly interesting candidate for a ring EDM experiment
and a proposal with a sensitivity of $10^{-27}$~e\,cm exists \cite{Semertzidis_2004}.
In this case scattering off a target will be used to observe a spin precession.
As possible sites of an experiment the Brookhaven National Laboratory (BNL),
the Indiana University Cyclotron Facility (IUCF) and the Kernfysisch Versneller
Instituut (KVI) are considered.

\begin{table}[bt]
\centering
\caption{\it Actual limits on permanent electric dipole moments.} 
{
\begin{tabular}{|c|c|c|} \hline

 Particle & Limit/Measurement [e-cm] & reference \\   \hline
e               & $<1.6 \times 10^{-27}$  & \cite{Regan} \\
$\mu$           & $<2.8\times 10^{-19}$ &  \cite{BNL_EDM} \\ 
$\tau$          & $(-2.2 <d_{\tau}<4.5)\times 10^{-17}$ &  \cite{Belle_tau} \\ 
n               & $<6.3 \times 10^{-26}$ &  \cite{Harris}  \\ 
p               & $(-3.7\pm 6.3) \times 10^{-23}$ & \cite{Proton_EDM}  \\ 
$\Lambda$       & $(-3.0\pm 7.4) \times 10^{-17}$ &  \cite{Lambda_EDM} \\ 
$\nu_{e,\mu}$   & $<2 \times 10^{-21}$ &  \cite{delAguila} \\ 
$\nu_{\tau}$    & $<5.2 \times 10^{-17}$ & \cite{NuTau_EDM}  \\   
Hg-atom         & $< 2.1 \times 10^{-28}$ & \cite{Romalis_2001}\\ \hline

\end{tabular}
}
\label{EDMs} 
\end{table}

\subsubsection{Correlations in $\beta$-decays}
In standard theory the structure of weak
interactions is V-A, which means there are vector (V) and axial-vector (A) 
currents with opposite relative sign causing a left handed structure 
of the interaction and parity violation \cite{Herczeg_2001}. 
Other possibilities like scalar, pseudo-scalar and tensor type 
interactions which might be possible would be clear 
signatures of new physics. So far they have been searched 
for without positive result. However, the bounds on parameters
are not very tight and leave room for various speculative possibilities. 
The double differential decay probability 
$ d^2W/d\Omega_e d\Omega_{\nu}$for a $\beta$-radioactive nucleus is
related to the electron and neutrino momenta $\vec{p}$ and $\vec{q}$ through
\begin{eqnarray}
\label{diffprob}
\frac{d^2W}{d\Omega_e d\Omega_{\nu}} & \sim & 1 +  a ~\frac{\vec{p}\cdot\vec{q}}{E} 
+  b ~~\sqrt{1-(Z \alpha)^2}~~\frac{m_e}{E}       \nonumber     \\
& & + <\vec{J}>     \cdot \left[ A~~ \frac{\vec{p}}{E} + B~~\vec{q} + D~~\frac{\vec{p} \times ~\vec{q}}{E} \right]\\
& &+ <\vec{\sigma}> \cdot \left[ G~~ \frac{\vec{p}}{E} + Q~~\vec{J} + R~~ <\vec{J}> \times ~\frac{\vec{q}}{E} \right] \nonumber
\end{eqnarray}
where   $m_e$ is the $\beta$-particle mass,
        $E$ its energy,
        $\vec{\sigma}$ its spin,  and
        $\vec{J}$ is the spin of the decaying nucleus.
        The coefficients D and R are studied in a number of
        experiments at this time and they are T violating in nature. 
        Here D is of particular interest for
further restricting model parameters. It describes the correlation between 
the neutrino and $\beta$-particle momentum vectors for spin polarized nuclei. 
The coefficient R is highly sensitive within a smaller set of 
speculative models, since in this region there exist some already well established
constraints, e.g., from searches for permanent electric dipole moments  \cite{Herczeg_2001}. 

From the experimental point of view, 
an efficient direct measurement of the neutrino momentum is not possible.
The recoiling nucleus can be detected instead and the neutrino 
momentum can be reconstructed using the  kinematics of the process.
Since the recoil nuclei have typical energies in the few 10 eV range,
precise measurements can only be performed, if the decaying isotopes are 
suspended using extreme shallow potential wells. Such exist, for example,
in atom traps formed by laser light, where many atomic species can be stored at 
temperatures below 1 mK. An overview over actual activities
can be found in \cite{Behr_2004}. 

Such research is being performed at a number of laboratories worldwide.
At KVI a new facility
is being set up, in which T-violation research will be a central scientific issue
\cite{Jungmann_2002,Turkstra_2002}. At this new facility
the isotopes of primary interest are $^{20}$Na,$^{21}$Na, $^{18}$Ne and $^{19}$Ne.
These atoms have suitable spectral lines for 
optical trapping and the nuclear properties allow to observe 
rather clean transitions.

A recent measurement at Berkeley, USA, the asymmetry parameter $a$ in the 
$\beta$-decay of  $^{21}$Na has been measured in optically trapped atoms
\cite{Scielzo_2004}. The value
differs from the present SM value by about 3 standard deviations.
Whether this is an indication of new physics 
reflected in new interactions in $\beta$-decay, this depends strongly
on the $\beta / (\beta +\gamma)$ decay branching ratio  for which
some 5 measurements exists which in part disagree significantly \cite{Endt_1990} 
New measurements are needed.
The most stringent limit on scalar interactions for $\beta$-neutrino correlation 
measurements comes from an experiment on the pure Fermi decay of $^{38m}$K at TRIUMF, where 
$a$ was extracted to 0.5~\% accuracy and in good agreement with
standard theory \cite{Gorelov_2004}.

\section{Properties of Known Basic Interactions}

\subsection{Electromagnetism and Fundamental Constants}

In the electro-weak part of the SM
very high  precision can be achieved for calculations,
in particular within Quantum Electrodynamics (QED), which is
the best tested field theory we know and a key element of the SM.
QED allows for extracting  accurate values of important fundamental 
constants from high precision experiments on free particles 
and light bound systems, where perturbative approaches work
very well for their theoretical description. Examples are the
fine structure constant $\alpha$ or the Rydberg constant R$_{\infty}$.
The obtained numbers are needed to describe the known 
interactions precisely. 
Furthermore, accurate calculations 
provide a basis to searches for deviations from SM predictions.
Such differences would reveal clear and undisputed signs of New Physics
and hints for the validity of speculative extensions to the SM.
For bound systems containing nuclei with high electric charges 
QED resembles a field theory with strong coupling and new 
theoretical methods are needed.

\subsubsection{Muonium}
The interpretation of measurements in the 
muonium atom, the bound
state of a $\mu^+$ and an $e^-$, is free of difficulties
arising from the structure of its constituents \cite{Jungmann_2004}. 
Thus QED predictions with two orders of 
magnitude higher accuracy than for the hydrogen atom are 
possible.
The ground state hyperfine splitting as 
well as the $1s-2s$ energy difference have been precisely determined 
recently. These measurements 
can be interpreted as QED tests or alternatively  
-assuming the validity of QED- 
as independent  measurements  of $\alpha$ as well as of muon properties 
(muon mass $m_{\mu}$ and muon magnetic moment $\mu_{\mu}$).
These experiments are statistics limited. Significantly
improved values would be possible at new intense muon sources.
There is a close connection between muonium spectroscopy
and a measurement of the muon magnetic anomaly $a_{\mu}$,
the relative deviation of the muon g-factor from the Dirac value 2.
Muonium spectroscopy provides precise values for  mass,
electric charge and magnetic moment of the muon.

\subsubsection{Muon Magnetic Anomaly}
Precise values of these fundamental constants are indispensable 
for the evaluation of the experimental results of
a muon g-2 measurement series 
in a magnetic storage ring 
at BNL \cite{Bennett_2004}.
The quantity $a_{\mu}$
arises from quantum effects and is mostly due to QED.
Further, there is a contribution from strong interactions of 58~ppm 
which arises from had- ronic vacuum polarization. 
The influence of weak interactions amounts to 1.3 ppm. 
Whereas QED and weak effects can be calculated from first principles,
the hadronic contribution
needs to be evaluated through a dispersion relation
and experimental input from $e^+$-$e^-$ annihilation into hardrons.
Up to now the relevant cross section was determined in the 
 essential energy region in the CMD experiment in Novosibirsk, Russia,
 or extracted from hadronic  $\tau$-decays measured in several setups.  
Calculations of the hadronic
part in $a_{\mu}$ depend on the choice of presently available 
experimental hadronic data and are obtained from an integration
over all energies. The results for $a_{\mu}$
differ by 3.0
respectively 1.6 standard deviations from the averaged experimental value.
Intense theoretical and experimental efforts are needed to solve 
the hadronic correction puzzle. Evaluations of the hadronic corrections
based on available new data 
on $e^+$-$e^-$ annihilation from the KLOE experiment in Frascati, Italy, appear to confirm earlier
values\cite{Aloisio_2004}, although in small energy intervals
significant differences exist in the cross sections from the different experiments.
For the muon magnetic anomaly improvements both in theory and
experiment are required, before a definite conclusion can be drawn
whether a hint of physics beyond standard theory \cite{Chavez_2004}  
has been seen. A continuation of the g-2 experiment with improved 
equipment and beams
was scientifically approved in 2004.

 \section{New Instrumentation Needed}

Progress in the field of low energy experiments to verify and test the SM
and to search for extensions to it would benefit in many cases 
significantly from new instrumentation and a new generation of particle sources. 
In particular, a high power proton driver would boost a large number of 
possible experiments which all have  a high and robust  discovery potential \cite{NUPECC_2004}.
In \cite{Jungmann_2004} two possible scenarios for a 1~GeV
and a 30~GeV machine are compared with respect to the physics prospects
and the needs of in part novel experimental approaches (see e.g. \cite{Aysto_2001}). 
Only a few, but important
experiments (like muon g-2) would definitely require the high energy beams. 
The availability of such a new facility would be desirable for a number 
of other fields as well, such as neutron scattering, in particular ultra-cold neutron research,
or a new ISOL facility (e.g. EURISOL)
for nuclear physics with nuclei far off the valley of stability. A joint
effort of several communities 
could benefit from synergy effects. Possibilities for such a machine 
could arise at CERN \cite{Aysto_2001,CERN_PD_2004}, FEMILAB, J-PARC and GSI
with either a high power linac or a true rapid cycling synchrotron.

\section{Conclusions}

 Nuclear physics and nuclear techniques offer a variety of possibilities to investigate fundamental 
 symmetries in phy- sics and to search for physics beyond the SM. 
 Experiments at Nuclear Physics facilities at low and 
intermediate energies 
offer in this 
respect a variety of possibilities which are complementary 
to approaches in High Energy physics and in some cases
exceed those significantly in their potential to steer
physical model building.

 The advantage of high particle fluxes at a Multi-Mega- watt facility allow
 higher sensitivity to rare processes because of higher statistics 
 and because also in part novel experimental approaches are enabled by the combination of
 particle number and an appropriate time structure of the beam. The field is looking forward
 to a rich future.

\begin{acknowledgement} 
The author would like to the members of the  NuPECC
Long Range Plan 2004 Fundamental Interaction working group \cite{NUPECC_2004}
for numerous fruitful discussions.
This work was supported in part by the Dutch Stichting
voor Fundamenteel Onderzoek der Materie (FOM) in the
framework of the TRI$\mu$P programme and by the European Community
through the NIPNET RTD project.
\end{acknowledgement}

%

\begin{thebibliography}{}

\bibitem{Lee_56}           T.D. Lee and  C.N. Yang, Phys. Rev. {\bf 98}, 1501  (1955)
\bibitem{NUPECC_2004}      K. Jungmann et al., in: ''NuPECC Long Range Plan 2004'',  
                           M.N. Harakeh et al. (eds.), (2004) 
\bibitem{neutrino_reviews} for a review see, e.g.:Y. Grossmann,hep-ph/0305245 (2003)
\bibitem{Gorham_2002}      P.Gorham et al., 
                           Nucl.Instrum.Meth. {\bf A490} 476 (2002)
\bibitem{deMeijer_2004}    R.G. de Meijer et al., Nuclear Physics News {\bf 14}(2), 20 (2004)
\bibitem{MainzTroitzk}     V.M. Lobashov et al., Phys. Lett. {\bf B460}, 227 (1999);
                           Ch. Kraus et al. hep-ex/0412056 (2004) 
\bibitem{Weinheimer_2003}  Ch. Weinheimer, hep-ex/0306057 (2003) and Nucl. Phys. {\bf B 118}, 279 (2003)
\bibitem{Osipowicz_2001}   A. Osipowicz et al, hep-ex/0109033 (2001) 
\bibitem{Klapdor_2004}     H.V. Klapdor-Kleingrothaus, Phys. Lett. {\bf B586}, 198 (2004)
\bibitem{Elliot_2004}      for a more detailed review see, e.g.: S.R. Elliott and J. Engel, 
                           J.Phys. {\bf G30}, R183 (2004)
\bibitem{PDG}              Particle Data Group, K. Hagiwara et al., Phys. Rev. {\bf D 66}, 010001 (2002); 
                           see also: 
                           S. Eidelmann et al., Phys. Lett. {\bf B 592}, 1 (2004)
\bibitem{Toner_2003}       I.S. Toner and J.C. Hardy, J.Phys. {\bf G29}, 197 (2003)
\bibitem{Abele_2004}       H. Abele et al., Eur. Phys. J. {\bf C 33}, 1 (2004) 
\bibitem{Czarnecki_2004}   A. Czarnecki, W. Marciano and A. Sirlin, hep/ph-0406324 (2004) 
                           and references therein 
\bibitem{Ellis_2004}       J. Ellis, hep/ph-0409360 (2004)
\bibitem{Hardy_2004}       J. Hardy, priv. com. (2004); see also: J. Hardy, this volume 
\bibitem{Aysto_2001}       J. \"Ayst\"o et al, hep-ph/0109217 (2001) 
\bibitem{Chung_2002}       J. Chung, Phys. Rev. {\bf D66}, 032004 (2002)
\bibitem{Godzev_1994}      A.A. Godzev et al., Phys. Lett {\bf B 338}, 212 (1994)
\bibitem{Czarnecki_1998}   A. Czarnecki and W. Marciano, Int.J.Mod.Phys. {\bf A13}, 
                           2235 (1998)  and references therein
\bibitem{Qweak}            D. Armstrong et al, proposal E02-020 to Jefferson Lab (2002)
\bibitem{Fortson}          N. Fortson, in: ''Parity Violation in Atoms and Polarized Electron
                           Scattering'', B.F. Bouchiat and M.A. Bouchiat (eds.), World
                           Scientific, Singapore, p. 244 (1999) 
\bibitem{Atutov_2003}      S.N. Atutov et al., Hyperfine Interactions {\bf 146-147}, 83 (2003)
\bibitem{Gomez_2004}       E. Gomez et al., physics/0412073 (2004)
\bibitem{Sakharov_1967}    A. Sakharov,JETP {\bf 5}, 32 (1967); M. Trodden, 
                           Rev. Mod. Phys. {\bf 71}, 1463 (1999) 
\bibitem{Ramsey_1999}      N. Ramsey, at ''Breit Symposium'', Yale (1999)                             
\bibitem{Sandars_2001}     P.G.H. Sandars, Contemp. Phys. {\bf 42}, 97 (2001) 
\bibitem{Dzuba_2001}       V. Dzuba et al., Phys. Rev. A {\bf 63}, 062101 (2001)
\bibitem{Engel_2004}       J. Engel, this volume; J. Engel et al.,Phys. Rev. {\bf C 68}, 025501 {2003}
\bibitem{Regan}            B.C. Regan et. al., Phys. Rev. Lett., {\bf 88},  071805(2002)
\bibitem{BNL_EDM}          R. McNabb et al., hep-ex/0407008 (2004) 
\bibitem{Belle_tau}        K. Inami et. al., Phys. Lett. B, {\bf 551}, 16 (2002)
\bibitem{Harris}           P. G. Harris et. al., Phys. Rev. Lett, {\bf 82}, 904 (1999)
\bibitem{Proton_EDM}       D. Cho et. al., Phys. Rev. Lett., {\bf 63}, 2559 (1989)
\bibitem{Lambda_EDM}       L. Pondrom et. al., Phys. Rev. D, {\bf 23}, 814 (1981)
\bibitem{delAguila}        F. del Aguila and M. Sher, Phys. Lett. B, {\bf 252}, 116 (1990)
\bibitem{NuTau_EDM}        R. Escribano and E. Masso, Phys. Lett. B, {\bf 395}, 369 (1997)
\bibitem{Romalis_2001}     M. Romalis, W. Griffith and N. Fortson,Phys.Rev.Lett. {\bf 86}, 
                           2505 (2001) 
\bibitem{Kawall_2001}      D. Kawall et al., Phys.Rev.Lett.{\bf 92}, 13307 (2004) 
\bibitem{Yannis_2003}      Y. Semertzidis et al., J-PARC Letter of Intent L22 (2003)
\bibitem{Farley_2004}      F.J.M. Farley et al., Phys. Rev. Lett. {\bf 93}, 052001 (2004)
\bibitem{Feng_2004}        W.-F. Chang and J.N. Ng, hep-ex/0307006 (2004)
\bibitem{Babu_2000}        K.S. Babu, B. Dutta and R. Mohapatra, Phys.Rev.Lett. {\bf 85} 5064 (2000);
                           B. Dutta and R. Mohapatra, Phys. Rev. {\bf D68} (2003) 113008
\bibitem{Semertzidis_2004} Y. Semertzidis et al., AIP Conf. Proc. {\bf 698}, 200 (2004)
\bibitem{Hisano_2004}      J. Hisano, hep-ph/0410038 (2004)
\bibitem{Liu_2004}         C.P Liu and R.G.E. Timmermans, nucl-th/0408060 (2004)
\bibitem{Herczeg_2001}     P. Herczeg, Prog. Part. and Nucl. Phys. {\bf 46}, 413 (2001)
\bibitem{Scielzo_2004}     N.D. Scielzo, et al., Phys. Rev. Lett. {\bf 93}, 102501 (2004)
\bibitem{Endt_1990}        P.M. Endt, Nucl.Phys. {\bf A521}, 1 (1990)
\bibitem{Gorelov_2004}     A. Gorelov et al., nucle-ex/041232 (2004)
\bibitem{Behr_2004}        J. Behr, this volume 
\bibitem{Jungmann_2002}    K. Jungmann, Acta Phys. Pol., {\bf 33} 2049 (2002)
\bibitem{Turkstra_2002}    J.W. Turkstra et al, Hyp. Interact. {\bf 127}, 533 (2000) 
\bibitem{Farley_2004}      F. Farley et al., Phys. Rev. Lett. {\bf 93}, 052001 (2004)  
\bibitem{Jungmann_2004}    K. Jungmann, nucl-ex/0404013 (2004) 
\bibitem{Bennett_2004}     G.W. Bennett et al., Phys. Rev. Lett.{\bf 92}, 168102 (2004)
                           see also: Phys. Rev. Lett. {\bf 89},101804 (2002) and 
                           H.N. Brown et al., Phys.Rev.Lett.{\bf 86}, 2227 (2001)
\bibitem{Aloisio_2004}     A. Aloisio et al.,hep-ex/0407048  (2004)
\bibitem{Chavez_2004}      H. Chavez and C.N. Ferreira, hep-ph/0410373 (2004) and references therein
\bibitem{Jungmann_2004}    K. Jungmann, Proceedings of INPC04 (in print)
\bibitem{Nesvizhevsky_2003}V.V. Nesvizhevsky et al., Phys. Rev. {\bf D 67}, 102002 (2003);
\bibitem{CERN_PD_2004}     A. Blondel et al., ''Physics with a Multi-MW Proton Source'', 
                           CERN-SPSC-2004-024 (2004) 
\end{thebibliography}
%

\end{document}